\documentclass[iop,apj]{emulateapj}
\usepackage{amsmath,amssymb,amstext}

\usepackage[breaklinks,colorlinks,citecolor=blue,linkcolor=magenta]{hyperref} 

\usepackage[all]{hypcap} 

\usepackage{aas_macros}
\usepackage{natbib}

\def \lcdm {$\Lambda$CDM}

\def \Msun{\ {\rm M_\odot}}

\def \Mpch{\ h^{-1}{\rm Mpc}}
\def \kpch{\ h^{-1}{\rm kpc}}

\def \aap  {A\&A}

\shorttitle{ Matter in the Beam \& temperature of dark matter}
\shortauthors{Mahdi et al.}

\begin{document}

\title{Matter in the beam: Weak lensing, substructures and the temperature of dark matter}
\author{Hareth S. Mahdi$^{1,2}$, Pascal J. Elahi$^{1}$, Geraint F. Lewis$^{1}$ and Chris Power$^{3}$}
\email{hareth@physics.usyd.edu.au}

\altaffiltext{1}{Sydney Institute for Astronomy, School of Physics, A28, The University of Sydney, NSW 2006, Australia}
\altaffiltext{2}{Department of Astronomy, University of Baghdad, Jadiryah, Baghdad 10071, Iraq}
\altaffiltext{3}{International Centre for Radio Astronomy Research, University of Western Australia, 35 Stirling Highway, Crawley, WA 6009, Australia}

\begin{abstract}
Warm Dark Matter (WDM) models offer an attractive alternative to the current Cold Dark Matter (CDM) cosmological model. We present a novel method to differentiate between WDM and CDM cosmologies, namely using weak lensing; this provides a unique probe as it is sensitive to all the ``matter in the beam'', not just dark matter haloes and the galaxies that reside in them, but also the diffuse material between haloes. We compare the weak lensing maps of CDM clusters to those in a WDM model corresponding to a thermally produced $0.5$~keV dark matter particle. Our analysis clearly shows that the weak lensing magnification, convergence and shear distributions can be used to distinguish between CDM and WDM models. WDM models {\em increase} the probability of weak magnifications, with the differences being significant to $\gtrsim5\sigma$, while leaving no significant imprint on the shear distribution. WDM clusters analysed in this work are more homogeneous than CDM ones, and the fractional decrease in the amount of material in haloes is proportional to the average increase in the magnification. This difference arises from matter that would be bound in compact haloes in CDM being smoothly distributed over much larger volumes at lower densities in WDM. Moreover, the signature does not solely lie in the probability distribution function but in the full spatial distribution of the convergence field. 
\end{abstract}

\keywords{Gravitational lensing: weak --- Galaxies: clusters --- Dark matter --- Cosmology: theory --- Methods: numerical.}
\maketitle

\section{Introduction}
\label{sec:intro}
Tensions between observations and predictions from the Cold Dark Matter cosmological model has renewed interest in other types of dark matter \cite[e.g.][]{finkbeiner2011,lovell2012,libeskind2013,vogelsberger2012c,vogelsberger2013b,maccio2013,baldi2012a,carlesi2014a,carlesi2014b}. One particular flavour of great interest is Warm Dark Matter (WDM), where the fundamental DM particle has an appreciable velocity at early times. One of the best-known examples is a sterile neutrino, which could explain observed neutrino oscillation rates and baryogenesis \cite[$\nu$MS; e.g.][]{asaka2006}. WDM models are not only attractive due to their ability to address issues in particle physics but their effect on cosmological structure formation. The non-negligible motions leaves a specific feature in the initial density perturbations from which dark matter haloes arise, namely a suppression of power in the density field below the so-called free-streaming scale. 

\par 
As a consequence, WDM models predict negligible numbers of haloes below the free-streaming scale \cite[][]{boylankolchin2012,schneider2012,libeskind2013,lovell2013}. This suppression may reconcile the difference between the observed and predicted number of low mass dwarf galaxies around large galaxies such as our own Milky Way, solving the so-called missing satellites problem \cite[e.g.][]{klypin1999, 1999ApJ...524L..19M}. For instance, \cite{lovell2012} showed that the resonantly produced sterile neutrino WDM models, with particle masses of $\sim2$~keV compatible with the Lyman-$\alpha$ bounds \cite[][]{boyarsky2009a,boyarsky2009b}, decrease the number of substructures residing in a MW-size halo, significantly alleviating the missing satellite problem. Although the simple suppression of power cannot alone account for all the small-scale discrepancies, \cite[e.g.][]{schneider2013b}, this family of models have the advantage of possibly explaining several other observational anomalies, such as $3.5$~keV X-ray line seen in clusters \cite[e.g.][]{bulbul2014a}. 

\par
Although most studies have focused on the effect of WDM on small-scale cosmic structures, the absence of small-scale power leaves its fingerprints on scales much larger than the free-streaming scale \cite[e.g.][]{obreschkow2013a}. Signatures of WDM at larger scales have generally been neglected since one of the desired features of WDM models is that they have the same large-scale matter distribution as CDM models while possibly resolving discrepancies on small scales. 

\par
Gravitational lensing is one such probe as it is sensitive to the entire underlying matter distribution \cite[e.g.][]{bartelmannschneider2001,schneider2003a,ellis2010}. In \cite{mahdi2014a} \& \cite{elahi2014a}, we showed that strong lensing by galaxy clusters can differentiate between WDM and CDM models. WDM clusters have larger Einstein radii and lensing cross sections. This result is contrary to the naive expectation that WDM clusters should have smaller strong lensing cross sections than their CDM counterparts as CDM clusters contain more subhaloes, and subhaloes increase the lensing cross section \cite[][]{xu2009}. This unexpected signature reduces the tension between observations and theory: observed clusters produce more lensed giant arcs than \lcdm\ predicts, the so-called arc statistics problem (\citealp{bartelmann1998}; see \citealp{meneghetti2013} for a review). WDM decreases this tension, although it does not fully resolve it.

\par
In this paper, we continue exploring novel approaches for probing signatures of WDM, specifically one that does not rely on rare alignments between galaxies required for strong lensing, namely weak lensing. Weak lensing is a powerful tool for probing our cosmology as it is sensitive not just to the density peaks corresponding to haloes but to all the matter in the beam. Historically it has been measured using the shear of background galaxies through the statistical correlation of observed galaxy ellipticities (e.g. \citealp{sheldon2009a}; \citealp{umetsu2014a}; and see \citealp{bartelmannschneider2001} for a review). The small weak-lensing magnification has primarily been detected through its effect on the number density of a flux-limited sample \cite[e.g.][]{hilderbrandt2009}, though more recent observations have made use of the effect on the observed magnitudes and sizes of background galaxies around lensing clusters \cite[e.g][]{schmidt2012a}, and the distortion in the shape of background galaxies \cite[e.g][]{heymans2012a,gruen2014a}. \cite{2015arXiv150701858G} showed that the weak magnification signal can be estimated by comparing the number density of galaxies in a patch of sky with the expected unlensed number density. 

\par 
Several techniques of cosmic shear measurements have been investigated by Shear TEsting Program (STEP \citealp{2006MNRAS.368.1323H, massey2007a}) and GRavitational lEnsing Accuracy Testing (GREAT \citealp{bridle2010a}). These studies highlight observational problems such as blurring, pixelisation and noise uncertainty, which must be taken into account in order to measure the shear with high accuracy. The problems are somewhat alleviated by stacking images, which increases the signal-to-noise ratio \citep{2009MNRAS.398..471L}. 

\par 
Furthermore, due to the small distortion caused by weak lensing, detecting the weak lensing signal with high signal-to-noise requires a large number of background galaxies. The Dark Energy Survey (DES) aims to reconstruct the cosmic shear using approximately 300 million galaxies by surveying an area 30 times larger than previous weak lensing surveys \citep{2005IJMPA..20.3121F}. This will eventually be supplanted by LSST, which will use billions of galaxies \citep{2009arXiv0912.0201L}. 

\par 
Here we demonstrate how the weak lensing magnification distribution provides a novel probe to measure the ``temperature'' of dark matter.  We first present our methods in Section \ref{sec:methods}, and our results in Section \ref{sec:results}. We conclude in Section \ref{sec:discussion}.

\section{Synthetic weak lensing measurements}
\label{sec:methods}
We study the weak lensing signature around galaxy clusters from a suite of cosmological zoom simulations using a ray tracing method. We stack the weak lensing from our clusters along multiple lines-of-sight and compare the magnifications in our two models.
\subsection{Simulations}
\label{sec:sims}
We use 10 pairs of clusters extracted from zoom simulations of $\Lambda$WDM and $\Lambda$CDM cosmologies. Here we briefly discuss our simulations, for further details see \cite{elahi2014a}. The cosmological parameters used were: $h=0.7$, $\Omega_m=0.3$, $\Omega_{\Lambda}= 0.7$, and $\sigma_8=0.9$. The WDM model used is a $0.5$~keV thermally produced dark matter particle \cite[][]{bode2001,power2013}, which results in a suppression of growth for halo with $M\lesssim M_{\rm hm}=2.1\times10^{11}\Msun$, the so-called half-mode mass scale where the WDM power spectrum is $1/4$ that of the CDM one \cite[][]{schneider2012}. Based on phase-space considerations, \cite{2013MNRAS.430.2346S} estimated the mass of the WDM particle to be $\sim0.5$~keV. WDM cosmologies with a dark matter mass of 0.5 keV have been explored by various studies \cite[e.g.][]{schneider2012,2012MNRAS.421...50V,2013ApJ...767...22K}. 

\par 
We note that the WDM initial conditions {\em do not} include non-gravitational velocities, thus technically, the WDM simulations are CDM ones with a smooth truncation in the initial density perturbation power spectrum at a scale corresponding to $0.75\Mpch$. All simulations were run with {\small GADGET2}, a TreePM code \citep{gadget2} and each pair of zoom simulations used the same gravitational softening length based on \cite{power2003}, ie: $\epsilon_{\rm opt}=4\,R_{\rm vir}/\sqrt{N_{\rm vir}}$ using $R_{\rm vir}$ from the parent CDM simulation. Clusters and their subhaloes are identified using {\sc VELOCIraptor} (aka {\sc STF}, \citealp{elahi2011}).

\subsection{Weak Lensing}
\label{sec:lensing}
\begin{figure}
\centering  
\includegraphics[width=0.99\columnwidth]{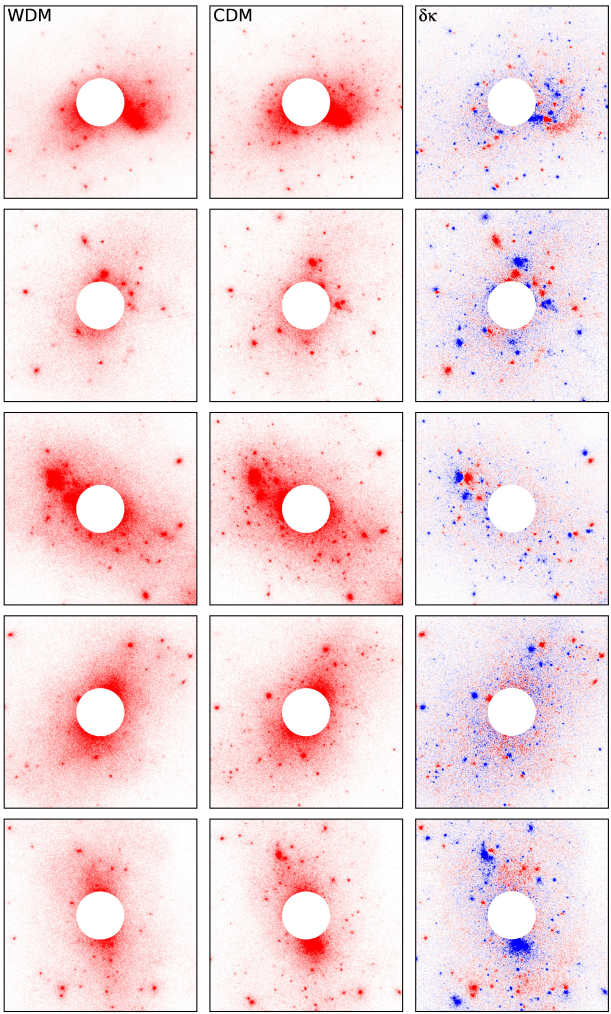}\vspace{-0.3cm}
\caption{Convergence maps of five WDM clusters (left column) and their CDM counterparts (middle column) as well as the difference between WDM \& CDM, $\delta\kappa\equiv\kappa_{\rm WDM}-\kappa_{\rm CDM}$ (right column). The side length of each map is 4 Mpc and the strong lensing regions (i.e. within 1 Mpc in diameter) are excluded. Here pixels with $\delta\kappa>0$ ($<0)$ are red (blue).}
\label{fig:kappa}
\end{figure}
Gravitational lensing probes the surface mass density of the lens $\Sigma$, specifically through the lensing potential $\psi$ along a line-of-sight (los): 
\begin{equation}\label{eq:potential}
    \psi({\bf \theta})=\frac{1}{\pi} \int \kappa({\bf \theta^{'}}) \ln |{\bf \theta}-{\bf \theta^{'}}| \,\mathrm{d}^{2} \theta^{'},
\end{equation}
where $\kappa=\Sigma/\Sigma_{\rm crit}$ is the convergence, $\Sigma_{\rm crit}= \frac{c^2}{4\pi G}\frac{D_{\rm S}}{D_{\rm L}D_{\rm LS}}$ is the critical surface mass density which depends on the angular distances to the source and lens, $D_{\rm L}$, $D_{\rm S}$ and between the lens and source $D_{\rm LS}$, and $\theta$ is the angle from the centre of the lens to the los. The magnification matrix $\mathcal{A}$ of an image can be written in terms of the convergence, and the two components of shear $\gamma_1$ and $\gamma_2$ as follows:
\begin{align}
  \mathcal{A}
      =\left(
      \begin{smallmatrix}
          1 - \frac{\partial^2 \psi}{\partial \theta_x^2 } & \frac{\partial^2 \psi}{\partial \theta_x \partial \theta_y} \\
          \frac{\partial^2 \psi}{\partial \theta_x \partial \theta_y} & 1 - \frac{\partial^2 \psi}{\partial \theta_y^2 }
      \end{smallmatrix}
      \right)
      =\left( 
      \begin{smallmatrix}
          1-\kappa-\gamma_1 &-\gamma_2 \\
          -\gamma_2 & 1-\kappa+\gamma_1
      \end{smallmatrix}
      \right)
\end{align}
The convergence gives the isotropic (de)magnification of an image due to the contribution of mass inside a bundle of light from a background source and the shear is responsible for the anisotropic distortion of images due to contribution of matter outside the bundle of light. 

\par 
The total magnification of an image is given by $\mu = 1/{\rm det}|\mathcal{A}| = [(1-\kappa)^2-\gamma^2]^{-1}$, where ${\bf \gamma}= \gamma_1 + i\gamma_2$ is the total shear. Note that for a simple azithumally symmetric lens, the equations simplify and one can show that $\gamma=[\bar\Sigma(<r)-\Sigma(r)]/\Sigma_{\rm cr}$, that is the shear probes the difference of the surface mass density. The weak lensing regime of a galaxy cluster takes place at large distance from the centre where ($\kappa \ll 1$ \& $\gamma \ll 1$). 

\par 
The weak lensing shear has a direct impact on the observed ellipticity of background sources. The observed ellipticity of a source ($e_{obs}$) is a combination of the source galaxy's intrinsic ellipticity ($e_s$) and the effect of shear,
\begin{equation}
e_{obs} = \frac{e_s+g}{1+g^*e_s}
\end{equation}    
where $g$ is the reduced shear, $g=\gamma/(1-\kappa)$, and $g^*$ is its complex conjugate. 
\par
We briefly outline our method for constructing weak lensing maps here, for a detailed description see \cite{mahdi2014a}. We place our lensing clusters at $z_{\rm L}=0.3$ and project all particles within a radius of $4$ Mpc of the cluster centre onto a 2D grid for 48 los to overcome the magnification bias. The projected density field is smoothed by a truncated Gaussian kernel with a smoothing scale of $5~\kpch$ in order to overcome the numerical noise due to the discreteness of N-body simulation. We use a $8192^2$ grid, resulting in an angular resolution of $0.22$~arcsec per pixel. 

\par
The convergence produced by five pairs of clusters along a single los as well as the difference $\delta\kappa\equiv\kappa_{\rm WDM}-\kappa_{\rm CDM}$ is shown in Figure \ref{fig:kappa}. To emphasise the differences in the weak lensing regime, we mask out the strong lensing region (1 Mpc in diameter). We see that in both cosmologies, the clusters contain small regions of high convergence, corresponding to the large subhaloes. As both the WDM and CDM simulations used the same phases in the initial conditions, every large subhalo in a CDM cluster has a counterpart in the WDM analogue. This is why large compact red regions $(\delta\kappa>0)$ are paired with a blue region $(\delta\kappa<0)$ in the difference map. The key feature is that more pixels with $\delta\kappa>0$ are smoothly distributed about the clusters. We find that $53\%$ of the surface area has $\delta\kappa>0$ for all clusters in our sample. We will discuss this feature in the following sections.

\par
For each projection, we distribute 30000 elliptical sources (galaxies) of random orientation and ellipticity on the source plane at $z_{\rm S}=2$ (i.e. $\sim$ 30 galaxies per square arcmin which is the effective number density practicable for weak lensing surveys \cite[see e.g.][]{2009arXiv0912.0201L}). We use the lens equation ${\bf \beta}={\bf \theta}-{\bf \alpha(\bf \theta)}$ to trace light rays of every single source from the source plane to the lens plane, identify the corresponding image(s) by means of a component-labelling algorithm that was proposed by \cite{Chang04alinear-time}. The convergence and shear of an identified image are calculated by averaging over all pixels that correspond to the image and the magnification is calculated from $\mu=1/|(1-\kappa)^2-\gamma^2|$.

\section{Results \& Analysis}
\label{sec:results}
We characterise the images of a multitude of sources produced by gravitational lenses, like those in Figure \ref{fig:kappa}. Naturally, outside the cluster centre there is little mass above the normal background density along a given los and background galaxies are not strongly magnified. We focus on the areas where the magnification is non-negligible and have ignored the areas where $1-\mu\lesssim10^{-3}$. We note that the form of the probability distribution function (PDF) is the same for all clusters.

\par 
The very outskirts of our clusters typically contain underdense regions. These void-like regions are poorly sampled even in our simulations resulting in areas with fewer than one particle per pixel. The convergence in these areas cannot be accurately measured and our method typically returns convergences of $\lesssim10^{-5}$. As WDM haloes are more extended \cite[][]{elahi2014a} and the voids in WDM cosmologies are not as underdense as those in CDM cosmologies \cite[][]{yang2015a}, the number of pixels with {\em very low} magnifications in our CDM simulations is artificially high. However, this does not affect our results.
\begin{figure}
  \centering
  \includegraphics[width=0.98\columnwidth]{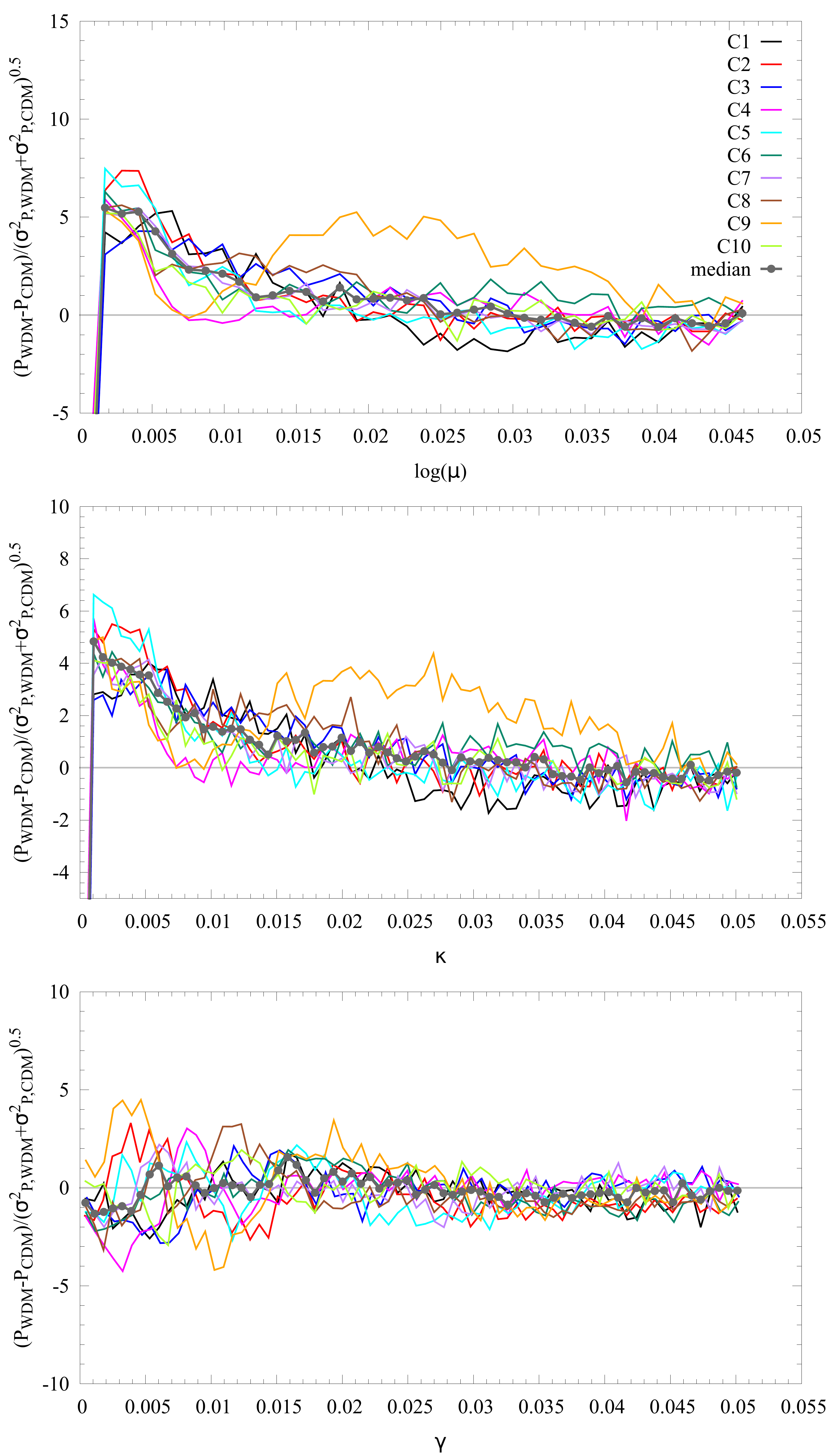}
  \caption[Caption for LOF]{Significance in the difference between WDM \& CDM PDFs of magnification (top), convergence (middle) and shear (bottom) for each cluster. The grey line represents the median distribution.}
  \label{fig:err}
\end{figure}

\par
To quantify the feature present in the convergence difference map seen in Figure \ref{fig:kappa}, we look at the differences in distribution of the convergence,magnification and shear of the identified images in the weak lensing regime (i.e. images with $\kappa,\gamma\leq 0.05$). We calculate the distribution of each quantity for each cluster separately and estimate the variation in the PDFs by bootstrap re-sampling using 100 sub-samples containing $1/5$ of the total number of the identified images \citep{NR}. Figure \ref{fig:err} shows the significance in the difference between the PDFs\footnotemark[1]. This figure shows that the convergence and magnification PDFs of WDM have an excess relative to CDM significant to $\gtrsim5\sigma$ for almost all clusters up to $\kappa\sim0.01$ and $\log\mu\sim0.01$. The one outlier is cluster 9 which shows a much larger excess in the WDM cosmology for large magnifications\footnotemark[2]. 
\footnotetext[1]{The first bin in this figure is affected by poorly sampled underdense regions where the convergence cannot be accurately measured and hence the difference in is not meaningful. Here it happens that there are more pixels with low convergence which are poorly measured in the CDM cosmology than that in the WDM one.}
\footnotetext[2]{This cluster does not appear to be very disturbed nor does it have a particularly unusual accretion history. However, there is a large group mass halo at the edge of the region used to calculate the weak lensing, which is closer to the main cluster in the WDM simulation. Additionally, the filamentary material bridging these objects is denser in the WDM simulation. }

\par
The difference in the shear distribution shows no significant features and the WDM and CDM PDFs are within $\lesssim2\sigma$. Again, cluster 9 appears to be an outlier from the rest of the clusters. 
\begin{figure}
    \centering  
    \includegraphics[width=0.99\columnwidth, trim=0.1cm 0.1cm 0.1cm 1.0cm, clip=true]{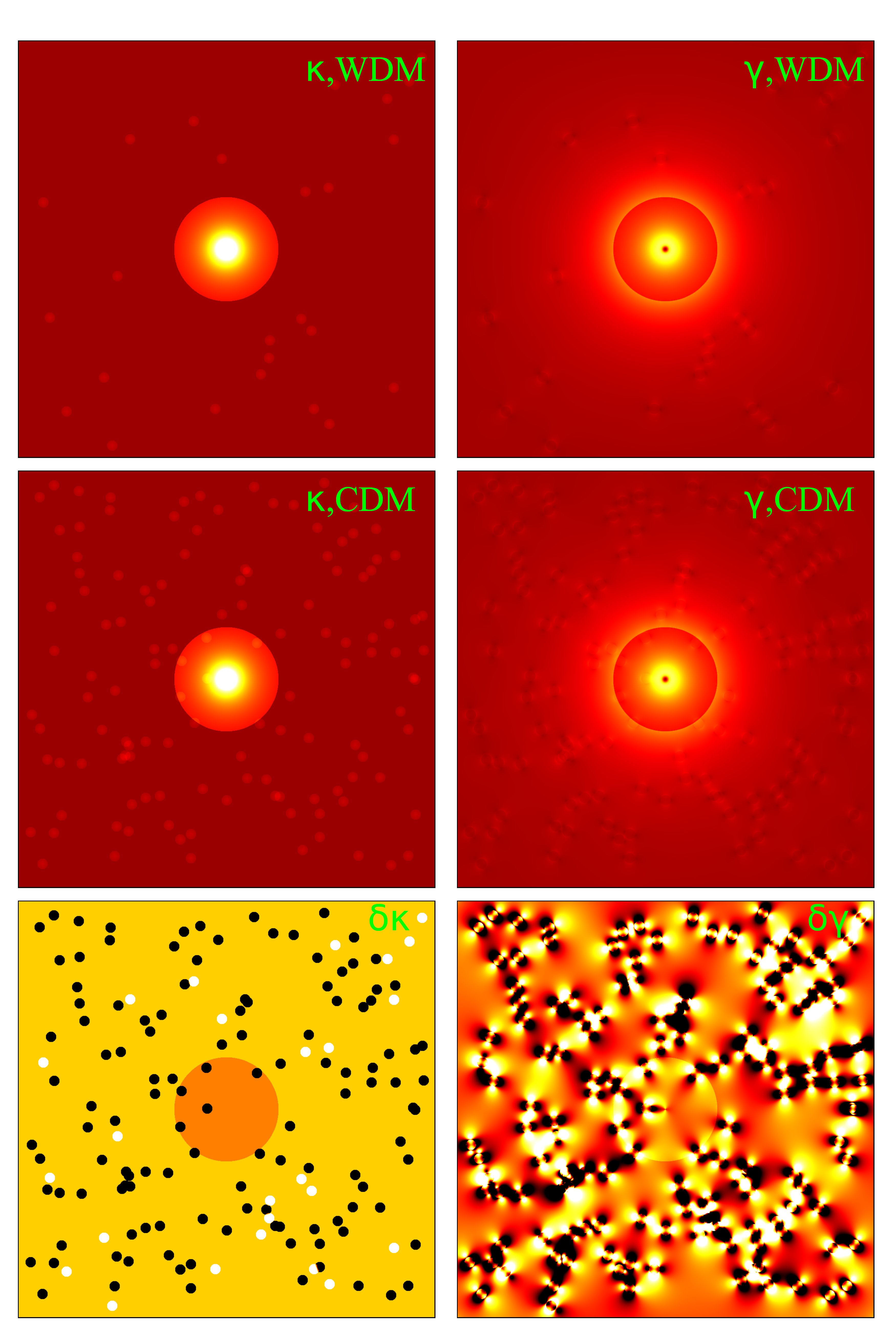}
    \caption{Lensing toy model of WDM \& CDM cosmologies. The left \& right columns show $\kappa$ \& $\gamma$ respectively. Top: WDM analogue, containing a single large halo and several smaller subhaloes. Middle: CDM analogue, which has more small subhaloes. Bottom: Differences in $\kappa$ \& $\gamma$. The colour scale in the two top rows are red $\sim0$ brightening to yellow for $\gg0$. Difference maps colours are yellow and white $=\delta\kappa,\delta\gamma>0$, orange $=\delta\kappa,\delta\gamma=0$, black $=\delta\kappa,\delta\gamma<0$.}
    \label{fig:toymodel}
\end{figure}

\par 
Where does this excess in the probability of weak lensing magnifications in WDM arise from? Figure \ref{fig:kappa} appeared to show a larger number of pixels with a positive $\delta\kappa$ between WDM and CDM and figure \ref{fig:err} also shows that the excess magnification arises solely from a higher convergence. As $\kappa$ is sensitive to the surface density and $\gamma$ probes the changes in the surface density, the difference in the models is {\em not} due to differences in the density profiles of subhaloes, nor necessarily subhaloes themselves. This higher convergence is quite simply due to there being more ``matter in the beam'' in regions giving rise to weak lensing. On the other hand, the fewer number of subhaloes in the WDM simulation results in fewer pixels with high shear when considering the whole shear field. However, as we focus on images in the weak lensing regime (i.e. with $\kappa,\gamma\leq 0.05$), the shear distribution shown in the lower panel of Figure \ref{fig:err} is not affected.

\par
The matter is distributed differently in WDM than it is in CDM, specifically, the material that would be bound up in compact low mass subhaloes is smoothly distributed over much larger volumes at lower densities. Put simply, WDM cosmologies are more homogeneous than CDM ones. 

\par 
One of the first discussion of how differences in the overall matter distribution affect observations was by \cite{dyerroeder1972,dyerroeder1973}, but have been considered in a number of subsequent publications \cite[e.g.][]{weinberg1976,nottale1982a,nottale1982b,nottale1983a,watanabe1990,kantowski1995,kantowski1998,tomita1999,kantowski2001,brouzakis2008,grenon2010a,clarkson2012,meures2012,fleury2013b}. The focus of these studies was to understand how the angular diameter distance depends upon the inhomogeneity of the universe, specifically on the fraction of material bound up in point masses along a los in comparison to a homogeneous background. The key finding from these studies is that inhomogeneities in the beam can provide additional (de)focusing, and hence distorting the observed relationship between redshift and the cosmological distance measures. Such inhomogeneities can have a significant influence and can even mimic the effects of dark energy on luminosity distances \cite[for a review see][]{bolejko2011}.

\par
A similar effect is at play here. Consider a simple toy model for a cluster in WDM \& CDM: in the WDM model, a volume contains a single large overdensity (halo), plus several less massive subhaloes randomly distributed within the volume and some smooth background; in the CDM one, the volume contains the same large halo along with a greater number of subhaloes randomly distributed within the volume. Here we treat haloes as truncated isothermal spheres, with the halo and subhaloes having truncation radii of $0.125L_{\rm box}$ and $0.0125L_{\rm box}$ respectively, where $L_{\rm box}$ is the size of the volume. For simplicity, we assume subhaloes have the same mass. We place down $N_{\rm sub,cdm}$ and $N_{\rm sub,wdm}$ subhaloes in the CDM and WDM volumes and set the total mass in subhaloes to be $M_{\rm sub,tot,cdm}=f_{\rm sub,cdm}M_{\rm tot}$, $M_{\rm sub,tot,wdm}=f_{\rm sub,wdm}M_{\rm tot}$, some fraction of the total mass in the CDM and WDM, respectively. As the mass within the volume is fixed and fewer subhaloes are present in the WDM, the background density in the CDM analogue is lower than that in the WDM analogue by
\begin{align}
\begin{split}
  \delta\rho_{\rm bg} &=(M_{\rm sub,tot,cdm}-M_{\rm sub,tot,wdm})/L_{\rm box}^3\\
  &=(f_{\rm sub,cdm}-f_{\rm sub,wdm})M_{\rm tot}/L_{\rm box}^3.
  \end{split}
\end{align}
The change in the background density manifests as a change in the convergence. This is seen in the $\delta\kappa$ panel of the toy model where most of the volume has $\delta\kappa>0$, though the regions where $\delta\kappa<0$ have large absolute changes in the convergence. 

\par
We illustrate these effects in Figure \ref{fig:toymodel}, where we plot the convergence and shear of the two toy models in the two upper rows and the difference in the bottom row. We have set $N_{\rm sub,cdm}=125$ \& $f_{\rm sub,cdm}=0.125$ in the CDM analogue and $N_{\rm sub,wdm}=25$ \& $f_{\rm sub,wdm}=0.025$ in the WDM analogue. Despite the fact that there are many more overdensities in the CDM volume, most the area has a higher convergence in the WDM toy model $(\delta\kappa>0)$. However, it is difficult to see if any bias is present in $\delta\gamma$. For this particular set of parameters we find $89\%$ of the surface area has $\delta\kappa>0$ compared to $6\%$ with $\delta\kappa<0$. The shear map is more evenly split, $42\%$ \& $58\%$ with $\delta\gamma>0$ \& $\delta\gamma<0$ respectively. Naturally the exact numbers depend not only on $f_{\rm sub}$ and $N_{\rm sub}$ but the size and distribution of subhaloes. Nevertheless, this model clearly shows that the higher fraction of material bound up in compact, high density regions in CDM cosmologies will produced smaller areas with non-negligible magnification compared to a WDM cosmology. WDM leaves the shear unchanged but increases the convergence.

\section{Discussion and Conclusions} 
\label{sec:discussion}
The removal of small-scale power in WDM models has far reaching effects. It is well known that it gives rise to fewer low mass dark matter haloes. However, the messy business of galaxy formation and evolution and the difficulty in finding small galaxies means that the galaxy luminosity function is a less than ideal probe for WDM. In this paper, we show that weak lensing offers another test for the absence of small-scale power. WDM models increase the average convergence (\& magnification) in the weak lensing regime, or more specifically the PDF distribution is shifted towards larger $\kappa$ for $\kappa\lesssim0.01$. Moreover, the signal is not only in the PDF from the entire lensing field. Key is that the fact that the spatial distributions of the convergence in the WDM and CDM cosmologies differ. It is visually apparent from the convergence maps (Figure \ref{fig:kappa}) that the matter in the WDM cosmology is more homogeneously distributed than that in the CDM one.    

\par 
The WDM model studied in this work assumes that the mass of warm dark matter particles is 0.5 keV. However, several other probes prefer higher values of WDM mass particles. Nevertheless, the key finding in this paper is that the matter is distributed differently in WDM models and in turn results in a difference in the weak lensing signature. Based on the results presented in this work, one would expect that the difference between the PDFs of the convergence and magnification to be smaller for WDM models with higher masses of  warm dark matter particles. A further work will consider studying the lensing characteristics of simulated clusters in WDM cosmologies with higher masses of WDM particles. This study will provide accurate statistical estimates of the difference between the CDM model and WDM cosmologies with different masses of WDM particles.    

\par 
The component-labelling algorithm used in this work exactly determines how the source has been distorted by the lens. The real world, however, is not so simple as we do not know the exact shape and intensity of the source so these lensing quantities are hidden. Real weak lensing observations probe the reduced shear by examining the shape of the flux distribution of a galaxy \cite[e.g.][]{sheldon2009a,umetsu2014a} or the magnification through the number density effect \cite[e.g.][]{hilderbrandt2009,ishigaki2015}. A future study will explore the detectability of WDM using mock weak lensing observations of galaxy shapes and the number density of galaxies for our sample of simulated clusters. Comparing these synthetic observations with measurements from the ongoing surveys such as DES and LSST would maximise the scientific impact of these measurements. 

\par 
In conclusion, this study highlights the power of weak lensing in using it as a precise cosmological probe. Weak lensing not only probes haloes but can measure the clumpiness of the intercluster medium. Therefore, in principle, it can be used to differentiate between cold and warm dark matter.

\acknowledgments{
  HSM is supported by the University of Sydney International Scholarship. PJE is supported by the SSimPL programme and the Sydney Institute for Astronomy (SIfA), DP130100117 and DP140100198. CP is supported by DP130100117, DP140100198, and FT130100041. GFL acknowledges financial support through DP130100117. This research was undertaken with the assistance of resources from the National Computational Infrastructure (NCI), which is supported by the Australian Government.
}

\bibliographystyle{apj}
\bibliography{paper}

\begin{thebibliography}{71}
\expandafter\ifx\csname natexlab\endcsname\relax\def\natexlab#1{#1}\fi

\bibitem[{{Asaka} {et~al}\mbox{.}(2006){Asaka}, {Shaposhnikov}, \&
  {Kusenko}}]{asaka2006}
{Asaka} T., {Shaposhnikov} M., {Kusenko} A., 2006, Physics Letters B, 638, 401

\bibitem[{{Baldi}(2012)}]{baldi2012a}
{Baldi} M., 2012, \mnras, 420, 430

\bibitem[{{Bartelmann} {et~al}\mbox{.}(1998){Bartelmann}, {Huss}, {Colberg},
  {Jenkins}, \& {Pearce}}]{bartelmann1998}
{Bartelmann} M., {Huss} A., {Colberg} J.~M., {Jenkins} A., {Pearce} F.~R.,
  1998, \aap, 330, 1

\bibitem[{{Bartelmann} \& {Schneider}(2001)}]{bartelmannschneider2001}
{Bartelmann} M., {Schneider} P., 2001, \physrep, 340, 291

\bibitem[{{Bode} {et~al}\mbox{.}(2001){Bode}, {Ostriker}, \&
  {Turok}}]{bode2001}
{Bode} P., {Ostriker} J.~P., {Turok} N., 2001, \apj, 556, 93

\bibitem[{{Bolejko} {et~al}\mbox{.}(2011){Bolejko}, {C{\'e}l{\'e}rier}, \&
  {Krasi{\'n}ski}}]{bolejko2011}
{Bolejko} K., {C{\'e}l{\'e}rier} M.-N., {Krasi{\'n}ski} A., 2011, Classical and
  Quantum Gravity, 28, 164002

\bibitem[{{Boyarsky} {et~al}\mbox{.}(2009{\natexlab{a}}){Boyarsky},
  {Lesgourgues}, {Ruchayskiy}, \& {Viel}}]{boyarsky2009a}
{Boyarsky} A., {Lesgourgues} J., {Ruchayskiy} O., {Viel} M.,
  2009{\natexlab{a}}, Physical Review Letters, 102, 201304

\bibitem[{{Boyarsky} {et~al}\mbox{.}(2009{\natexlab{b}}){Boyarsky},
  {Ruchayskiy}, \& {Shaposhnikov}}]{boyarsky2009b}
{Boyarsky} A., {Ruchayskiy} O., {Shaposhnikov} M., 2009{\natexlab{b}}, Annual
  Review of Nuclear and Particle Science, 59, 191

\bibitem[{{Boylan-Kolchin} {et~al}\mbox{.}(2012){Boylan-Kolchin}, {Bullock}, \&
  {Kaplinghat}}]{boylankolchin2012}
{Boylan-Kolchin} M., {Bullock} J.~S., {Kaplinghat} M., 2012, \mnras, 422, 1203

\bibitem[{{Bridle} {et~al}\mbox{.}(2010){Bridle}, {Balan}, {Bethge}, {Gentile},
  {Harmeling}, {Heymans}, {Hirsch}, {Hosseini}, {Jarvis}, {Kirk}, {Kitching},
  {Kuijken}, {Lewis}, {Paulin-Henriksson}, {Sch{\"o}lkopf}, {Velander},
  {Voigt}, {Witherick}, {Amara}, {Bernstein}, {Courbin}, {Gill}, {Heavens},
  {Mandelbaum}, {Massey}, {Moghaddam}, {Rassat}, {R{\'e}fr{\'e}gier}, {Rhodes},
  {Schrabback}, {Shawe-Taylor}, {Shmakova}, {van Waerbeke}, \&
  {Wittman}}]{bridle2010a}
{Bridle} S. {et~al.}, 2010, \mnras, 405, 2044

\bibitem[{{Brouzakis} {et~al}\mbox{.}(2008){Brouzakis}, {Tetradis}, \&
  {Tzavara}}]{brouzakis2008}
{Brouzakis} N., {Tetradis} N., {Tzavara} E., 2008, \jcap, 4, 8

\bibitem[{{Bulbul} {et~al}\mbox{.}(2014){Bulbul}, {Markevitch}, {Foster},
  {Smith}, {Loewenstein}, \& {Randall}}]{bulbul2014a}
{Bulbul} E., {Markevitch} M., {Foster} A., {Smith} R.~K., {Loewenstein} M.,
  {Randall} S.~W., 2014, \apj, 789, 13

\bibitem[{{Carlesi} {et~al}\mbox{.}(2014{\natexlab{a}}){Carlesi}, {Knebe},
  {Lewis}, {Wales}, \& {Yepes}}]{carlesi2014a}
{Carlesi} E., {Knebe} A., {Lewis} G.~F., {Wales} S., {Yepes} G.,
  2014{\natexlab{a}}, \mnras

\bibitem[{{Carlesi} {et~al}\mbox{.}(2014{\natexlab{b}}){Carlesi}, {Knebe},
  {Lewis}, \& {Yepes}}]{carlesi2014b}
{Carlesi} E., {Knebe} A., {Lewis} G.~F., {Yepes} G., 2014{\natexlab{b}}, \mnras

\bibitem[{Chang {et~al}\mbox{.}(2004)Chang, jen Chen, \& jen
  Lu}]{Chang04alinear-time}
Chang F., jen Chen C., jen Lu C., 2004, Computer Vision and Image
  Understanding, 93, 206

\bibitem[{{Clarkson} {et~al}\mbox{.}(2012){Clarkson}, {Ellis}, {Faltenbacher},
  {Maartens}, {Umeh}, \& {Uzan}}]{clarkson2012}
{Clarkson} C., {Ellis} G.~F.~R., {Faltenbacher} A., {Maartens} R., {Umeh} O.,
  {Uzan} J.-P., 2012, \mnras, 426, 1121

\bibitem[{{Dyer} \& {Roeder}(1972)}]{dyerroeder1972}
{Dyer} C.~C., {Roeder} R.~C., 1972, \apjl, 174, L115

\bibitem[{{Dyer} \& {Roeder}(1973)}]{dyerroeder1973}
{Dyer} C.~C., {Roeder} R.~C., 1973, \apjl, 180, L31

\bibitem[{{Elahi} {et~al}\mbox{.}(2014){Elahi}, {Mahdi}, {Power}, \&
  {Lewis}}]{elahi2014a}
{Elahi} P.~J., {Mahdi} H.~S., {Power} C., {Lewis} G.~F., 2014, \mnras, 444,
  2333

\bibitem[{{Elahi} {et~al}\mbox{.}(2011){Elahi}, {Thacker}, \&
  {Widrow}}]{elahi2011}
{Elahi} P.~J., {Thacker} R.~J., {Widrow} L.~M., 2011, \mnras, 418, 320

\bibitem[{{Ellis}(2010)}]{ellis2010}
{Ellis} R.~S., 2010, Royal Society of London Philosophical Transactions Series
  A, 368, 967

\bibitem[{{Finkbeiner} {et~al}\mbox{.}(2011){Finkbeiner}, {Goodenough},
  {Slatyer}, {Vogelsberger}, \& {Weiner}}]{finkbeiner2011}
{Finkbeiner} D.~P., {Goodenough} L., {Slatyer} T.~R., {Vogelsberger} M.,
  {Weiner} N., 2011, \jcap, 5, 2

\bibitem[{{Flaugher}(2005)}]{2005IJMPA..20.3121F}
{Flaugher} B., 2005, International Journal of Modern Physics A, 20, 3121

\bibitem[{{Fleury} {et~al}\mbox{.}(2013){Fleury}, {Dupuy}, \&
  {Uzan}}]{fleury2013b}
{Fleury} P., {Dupuy} H., {Uzan} J.-P., 2013, \prd, 87, 123526

\bibitem[{{Gillis} \& {Taylor}(2015)}]{2015arXiv150701858G}
{Gillis} B., {Taylor} A., 2015, ArXiv e-prints

\bibitem[{{Grenon} \& {Lake}(2010)}]{grenon2010a}
{Grenon} C., {Lake} K., 2010, \prd, 81, 023501

\bibitem[{{Gruen} {et~al}\mbox{.}(2014){Gruen}, {Seitz}, {Brimioulle},
  {Kosyra}, {Koppenhoefer}, {Lee}, {Bender}, {Riffeser}, {Eichner},
  {Weidinger}, \& {Bierschenk}}]{gruen2014a}
{Gruen} D. {et~al.}, 2014, \mnras, 442, 1507

\bibitem[{{Heymans} {et~al}\mbox{.}(2006){Heymans}, {Van Waerbeke}, {Bacon},
  {Berge}, {Bernstein}, {Bertin}, {Bridle}, {Brown}, {Clowe}, {Dahle}, {Erben},
  {Gray}, {Hetterscheidt}, {Hoekstra}, {Hudelot}, {Jarvis}, {Kuijken},
  {Margoniner}, {Massey}, {Mellier}, {Nakajima}, {Refregier}, {Rhodes},
  {Schrabback}, \& {Wittman}}]{2006MNRAS.368.1323H}
{Heymans} C. {et~al.}, 2006, \mnras, 368, 1323

\bibitem[{{Heymans} {et~al}\mbox{.}(2012){Heymans}, {Van Waerbeke}, {Miller},
  {Erben}, {Hildebrandt}, {Hoekstra}, {Kitching}, {Mellier}, {Simon},
  {Bonnett}, {Coupon}, {Fu}, {Harnois D{\'e}raps}, {Hudson}, {Kilbinger},
  {Kuijken}, {Rowe}, {Schrabback}, {Semboloni}, {van Uitert}, {Vafaei}, \&
  {Velander}}]{heymans2012a}
{Heymans} C. {et~al.}, 2012, \mnras, 427, 146

\bibitem[{{Hildebrandt} {et~al}\mbox{.}(2009){Hildebrandt}, {van Waerbeke}, \&
  {Erben}}]{hilderbrandt2009}
{Hildebrandt} H., {van Waerbeke} L., {Erben} T., 2009, \aap, 507, 683

\bibitem[{{Ishigaki} {et~al}\mbox{.}(2015){Ishigaki}, {Kawamata}, {Ouchi},
  {Oguri}, {Shimasaku}, \& {Ono}}]{ishigaki2015}
{Ishigaki} M., {Kawamata} R., {Ouchi} M., {Oguri} M., {Shimasaku} K., {Ono} Y.,
  2015, \apj, 799, 12

\bibitem[Kang et al.(2013)]{2013ApJ...767...22K} Kang, X., Macci{\`o}, 
A.~V., \& Dutton, A.~A.\ 2013, \apj, 767, 22 

\bibitem[{{Kantowski}(1998)}]{kantowski1998}
{Kantowski} R., 1998, \apj, 507, 483

\bibitem[{{Kantowski} \& {Thomas}(2001)}]{kantowski2001}
{Kantowski} R., {Thomas} R.~C., 2001, \apj, 561, 491

\bibitem[{{Kantowski} {et~al}\mbox{.}(1995){Kantowski}, {Vaughan}, \&
  {Branch}}]{kantowski1995}
{Kantowski} R., {Vaughan} T., {Branch} D., 1995, \apj, 447, 35

\bibitem[{{Klypin} {et~al}\mbox{.}(1999){Klypin}, {Gottl{\"o}ber}, {Kravtsov},
  \& {Khokhlov}}]{klypin1999}
{Klypin} A., {Gottl{\"o}ber} S., {Kravtsov} A.~V., {Khokhlov} A.~M., 1999,
  \apj, 516, 530

\bibitem[{{Lewis}(2009)}]{2009MNRAS.398..471L}
{Lewis} A., 2009, \mnras, 398, 471

\bibitem[{{Libeskind} {et~al}\mbox{.}(2013){Libeskind}, {Di Cintio}, {Knebe},
  {Yepes}, {Gottl{\"o}ber}, {Steinmetz}, {Hoffman}, \&
  {Martinez-Vaquero}}]{libeskind2013}
{Libeskind} N.~I., {Di Cintio} A., {Knebe} A., {Yepes} G., {Gottl{\"o}ber} S.,
  {Steinmetz} M., {Hoffman} Y., {Martinez-Vaquero} L.~A., 2013, Publications of
  the Astronomical Society of Australia, 30, 39

\bibitem[{{Lovell} {et~al}\mbox{.}(2012){Lovell}, {Eke}, {Frenk}, {Gao},
  {Jenkins}, {Theuns}, {Wang}, {White}, {Boyarsky}, \&
  {Ruchayskiy}}]{lovell2012}
{Lovell} M.~R. {et~al.}, 2012, \mnras, 420, 2318

\bibitem[{{Lovell} {et~al}\mbox{.}(2014){Lovell}, {Frenk}, {Eke}, {Jenkins},
  {Gao}, \& {Theuns}}]{lovell2013}
{Lovell} M.~R., {Frenk} C.~S., {Eke} V.~R., {Jenkins} A., {Gao} L., {Theuns}
  T., 2014, \mnras, 439, 300

\bibitem[{{LSST Science Collaboration}(2009){LSST Science
  Collaboration}, {Abell}, {Allison}, {Anderson}, {Andrew}, {Angel}, {Armus},
  {Arnett}, {Asztalos}, {Axelrod}, \& et~al.}]{2009arXiv0912.0201L}
{LSST Science Collaboration} {et~al.}, 2009, ArXiv e-prints

\bibitem[{{Macci{\`o}} {et~al}\mbox{.}(2013){Macci{\`o}}, {Ruchayskiy},
  {Boyarsky}, \& {Mu{\~n}oz-Cuartas}}]{maccio2013}
{Macci{\`o}} A.~V., {Ruchayskiy} O., {Boyarsky} A., {Mu{\~n}oz-Cuartas} J.~C.,
  2013, \mnras, 428, 882

\bibitem[{{Mahdi} {et~al}\mbox{.}(2014){Mahdi}, {van Beek}, {Elahi}, {Lewis},
  {Power}, \& {Killedar}}]{mahdi2014a}
{Mahdi} H.~S., {van Beek} M., {Elahi} P.~J., {Lewis} G.~F., {Power} C.,
  {Killedar} M., 2014, \mnras, 441, 1954

\bibitem[{{Massey} {et~al}\mbox{.}(2007){Massey}, {Heymans}, {Berg{\'e}},
  {Bernstein}, {Bridle}, {Clowe}, {Dahle}, {Ellis}, {Erben}, {Hetterscheidt},
  {High}, {Hirata}, {Hoekstra}, {Hudelot}, {Jarvis}, {Johnston}, {Kuijken},
  {Margoniner}, {Mandelbaum}, {Mellier}, {Nakajima}, {Paulin-Henriksson},
  {Peeples}, {Roat}, {Refregier}, {Rhodes}, {Schrabback}, {Schirmer}, {Seljak},
  {Semboloni}, \& {van Waerbeke}}]{massey2007a}
{Massey} R. {et~al.}, 2007, \mnras, 376, 13

\bibitem[{{Meneghetti} {et~al}\mbox{.}(2013){Meneghetti}, {Bartelmann},
  {Dahle}, \& {Limousin}}]{meneghetti2013}
{Meneghetti} M., {Bartelmann} M., {Dahle} H., {Limousin} M., 2013, \ssr, 177,
  31

\bibitem[{{Meures} \& {Bruni}(2012)}]{meures2012}
{Meures} N., {Bruni} M., 2012, \mnras, 419, 1937

\bibitem[{{Moore} {et~al}\mbox{.}(1999){Moore}, {Ghigna}, {Governato}, {Lake},
  {Quinn}, {Stadel}, \& {Tozzi}}]{1999ApJ...524L..19M}
{Moore} B., {Ghigna} S., {Governato} F., {Lake} G., {Quinn} T., {Stadel} J.,
  {Tozzi} P., 1999, \apjl, 524, L19
  
\bibitem[{{Nottale}(1982{\natexlab{a}})}]{nottale1982a}
{Nottale} L., 1982{\natexlab{a}}, \aap, 110, 9

\bibitem[{{Nottale}(1982{\natexlab{b}})}]{nottale1982b}
{Nottale} L., 1982{\natexlab{b}}, \aap, 114, 261

\bibitem[{{Nottale}(1983)}]{nottale1983a}
{Nottale} L., 1983, \aap, 118, 85

\bibitem[{{Obreschkow} {et~al}\mbox{.}(2013){Obreschkow}, {Power}, {Bruderer},
  \& {Bonvin}}]{obreschkow2013a}
{Obreschkow} D., {Power} C., {Bruderer} M., {Bonvin} C., 2013, \apj, 762, 115

\bibitem[{{Power}(2013)}]{power2013}
{Power} C., 2013, Publications of the Astronomical Society of Australia, 30, 53

\bibitem[{{Power} {et~al}\mbox{.}(2003){Power}, {Navarro}, {Jenkins}, {Frenk},
  {White}, {Springel}, {Stadel}, \& {Quinn}}]{power2003}
{Power} C., {Navarro} J.~F., {Jenkins} A., {Frenk} C.~S., {White} S.~D.~M.,
  {Springel} V., {Stadel} J., {Quinn} T., 2003, \mnras, 338, 14

\bibitem[{Press {et~al}\mbox{.}(2007)Press, Teukolsky, Vetterling, \&
  Flannery}]{NR}
Press W.~H., Teukolsky S.~A., Vetterling W.~T., Flannery B.~P., 2007, Numerical
  Recipes 3rd Edition: The Art of Scientific Computing, 3rd edn. Cambridge
  University Press, New York, NY, USA

\bibitem[Shao et al.(2013)]{2013MNRAS.430.2346S} Shao, S., Gao, L., Theuns, 
T., \& Frenk, C.~S.\ 2013, \mnras, 430, 2346 

\bibitem[{{Schmidt} {et~al}\mbox{.}(2012){Schmidt}, {Leauthaud}, {Massey},
  {Rhodes}, {George}, {Koekemoer}, {Finoguenov}, \& {Tanaka}}]{schmidt2012a}
{Schmidt} F., {Leauthaud} A., {Massey} R., {Rhodes} J., {George} M.~R.,
  {Koekemoer} A.~M., {Finoguenov} A., {Tanaka} M., 2012, \apjl, 744, L22

\bibitem[{{Schneider} {et~al}\mbox{.}(2014){Schneider}, {Anderhalden},
  {Macci{\`o}}, \& {Diemand}}]{schneider2013b}
{Schneider} A., {Anderhalden} D., {Macci{\`o}} A.~V., {Diemand} J., 2014,
  \mnras, 441, L6

\bibitem[{{Schneider} {et~al}\mbox{.}(2012){Schneider}, {Smith}, {Macci{\`o}},
  \& {Moore}}]{schneider2012}
{Schneider} A., {Smith} R.~E., {Macci{\`o}} A.~V., {Moore} B., 2012, \mnras,
  424, 684

\bibitem[{{Schneider}(2003)}]{schneider2003a}
{Schneider} P., 2003, ArXiv Astrophysics e-prints

\bibitem[{{Sheldon} {et~al}\mbox{.}(2009){Sheldon}, {Johnston}, {Scranton},
  {Koester}, {McKay}, {Oyaizu}, {Cunha}, {Lima}, {Lin}, {Frieman}, {Wechsler},
  {Annis}, {Mandelbaum}, {Bahcall}, \& {Fukugita}}]{sheldon2009a}
{Sheldon} E.~S. {et~al.}, 2009, \apj, 703, 2217

\bibitem[{{Springel}(2005)}]{gadget2}
{Springel} V., 2005, \mnras, 364, 1105

\bibitem[{{Tomita} {et~al}\mbox{.}(1999){Tomita}, {Asada}, \&
  {Hamana}}]{tomita1999}
{Tomita} K., {Asada} H., {Hamana} T., 1999, Progress of Theoretical Physics
  Supplement, 133, 155

\bibitem[{{Umetsu} {et~al}\mbox{.}(2014){Umetsu}, {Medezinski}, {Nonino},
  {Merten}, {Postman}, {Meneghetti}, {Donahue}, {Czakon}, {Molino}, {Seitz},
  {Gruen}, {Lemze}, {Balestra}, {Ben{\'{\i}}tez}, {Biviano}, {Broadhurst},
  {Ford}, {Grillo}, {Koekemoer}, {Melchior}, {Mercurio}, {Moustakas}, {Rosati},
  \& {Zitrin}}]{umetsu2014a}
{Umetsu} K. {et~al.}, 2014, \apj, 795, 163

\bibitem[Viel et al.(2012)]{2012MNRAS.421...50V} Viel, M., Markovi{\v c}, 
K., Baldi, M., \& Weller, J.\ 2012, \mnras, 421, 50 

\bibitem[{{Vogelsberger} \& {Zavala}(2013)}]{vogelsberger2013b}
{Vogelsberger} M., {Zavala} J., 2013, \mnras, 430, 1722

\bibitem[{{Vogelsberger} {et~al}\mbox{.}(2012){Vogelsberger}, {Zavala}, \&
  {Loeb}}]{vogelsberger2012c}
{Vogelsberger} M., {Zavala} J., {Loeb} A., 2012, \mnras, 423, 3740

\bibitem[{{Watanabe} \& {Tomita}(1990)}]{watanabe1990}
{Watanabe} K., {Tomita} K., 1990, \apj, 355, 1

\bibitem[{{Weinberg}(1976)}]{weinberg1976}
{Weinberg} S., 1976, \apjl, 208, L1

\bibitem[{{Xu} {et~al}\mbox{.}(2009){Xu}, {Mao}, {Wang}, {Springel}, {Gao},
  {White}, {Frenk}, {Jenkins}, {Li}, \& {Navarro}}]{xu2009}
{Xu} D.~D. {et~al.}, 2009, \mnras, 398, 1235

\bibitem[{{Yang} {et~al}\mbox{.}(2015){Yang}, {Neyrinck}, {Arag{\'o}n-Calvo},
  {Falck}, \& {Silk}}]{yang2015a}
{Yang} L.~F., {Neyrinck} M.~C., {Arag{\'o}n-Calvo} M.~A., {Falck} B., {Silk}
  J., 2015, \mnras, 451, 3606

\end{thebibliography}

\end{document}